\documentclass{aa}

\usepackage{txfonts}
\usepackage{longtable}
\usepackage{rotating}
\usepackage{natbib}
\usepackage{graphicx}
\usepackage{graphics}
\usepackage{psfrag}
\usepackage{amssymb}
\usepackage{xspace}
\usepackage{color}
\bibpunct{(}{)}{;}{a}{}{,}

\def\ne{n_{\rm e}}

\def\Teq{$T_{\rm eq}$}

\def\Rpl{$R_{\rm pl}$}
\def\Mpl{$M_{\rm pl}$}
\def\Re{\ensuremath{R_{\oplus}}}
\def\Me{\ensuremath{M_{\oplus}}}

\def\Msun{M$_{\odot}$}

\begin{document}
\title{The Kepler-11 system: evolution of the stellar high-energy emission and {initial planetary} atmospheric mass fractions}
\subtitle{}
\author{D. Kubyshkina\inst{1}   \and
    L. Fossati\inst{1}          \and
    A. J. Mustill\inst{2}       \and
    P. E. Cubillos\inst{1}      \and
    M. B. Davies\inst{2}            \and
    N. V. Erkaev\inst{3,4}      \and
    C. P. Johnstone\inst{5}     \and
    K. G. Kislyakova\inst{5,1}  \and
    H. Lammer\inst{1}           \and
    M. Lendl\inst{1,6}          \and
    P. Odert\inst{7}
}
\institute{
    Space Research Institute, Austrian Academy of Sciences, Schmiedlstrasse 6, A-8042 Graz, Austria\\
    \email{daria.kubyshkina@oeaw.ac.at}
    \and
    Lund Observatory, Department of Astronomy and Theoretical Physics, Lund University, Box 43, SE-221 00 Lund, Sweden
    \and
    Institute of Computational Modelling, SB RAS, 660036, Krasnoyarsk, Russia
    \and
    Siberian Federal University, 660041, Krasnoyarsk, Russian Federation
    \and
    Institute for Astronomy, University of Vienna, T\"urkenschanzstrasse 17,    A-1180 Vienna, Austria
    \and
    Observatoire astronomique de l'Universit\'e de Gen\`eve 51 ch. des Maillettes, 1290 Sauverny, Switzerland
    \and
    Institute of Physics/IGAM, University of Graz, Universit\"{a}tsplatz 5, A-8010 Graz, Austria
}
\date{}
\abstract {The atmospheres of close-in planets are strongly
influenced by mass loss driven by the high-energy (X-ray and
extreme ultraviolet, EUV) irradiation of the host star,
particularly during the early stages of evolution. We recently
developed a framework to exploit this connection and enable us to
recover the past evolution of the stellar high-energy emission
from the present-day properties of its planets, {if the latter
retain some remnants of their primordial hydrogen-dominated
atmospheres.} Furthermore, the framework can also provide
constraints on planetary initial atmospheric mass fractions. The
constraints on the output parameters improve when more planets can
be simultaneously analysed. This makes the Kepler-11 system, which
hosts six planets with bulk densities between 0.66 and
2.45\,g\,cm$^{-3}$, an ideal target. {Our results indicate} that
the star has likely evolved as a slow rotator (slower than 85\% of
the stars with similar masses), corresponding to a high-energy
emission at 150\,Myr of between 1-10 times that of the current
Sun. We also constrain the initial atmospheric mass fractions for
the planets, obtaining {a lower limit of 4.1\% for planet c, a
range of 3.7--5.3\% for planet d, a range of 11.1--14\% for planet
e, a range of 1--15.6\% for planet f, and a range of 4.7--8.7\%
for planet g assuming a disc dispersal time of 1\,Myr. For planet
b, the range remains poorly constrained.} Our framework also
suggests slightly higher masses for planets b, c, and f than have
been suggested based on transit timing variation measurements. We
coupled our results with published planet atmosphere accretion
models to obtain a temperature {(at 0.25 AU, the location of
planet f)} and dispersal time of the protoplanetary disc of 550\,K
and 1\,Myr,  although these results may be affected by
inconsistencies in the adopted system parameters. This work shows
that our framework is capable of constraining important properties
of planet formation models.}
\titlerunning{The Kepler-11 system}
\authorrunning{D. Kubyshkina et al.}
\maketitle
\section{Introduction}\label{sec::intro}

During the early stages of evolution, up to about 1--2\,Gyr,
late-type stars can follow different evolutionary paths in terms
of their rotation rate {and} hence high-energy
(X-ray\,$\text{plus}$\, extreme ultraviolet, EUV: hereafter XUV)
emission \citep[e.g.][]{tu2015,johnstone2015rot}. Furthermore, the
XUV emission from young stars has a decisive effect on the
atmospheric loss and evolution of their planets
\citep[e.g.][]{lopez2013,owen2018}. Therefore, the present-day
properties of planetary atmospheres are intimately related to the
evolutionary paths that were followed by their host stars. {In
general, hydrogen-rich planets are expected to orbit low-activity
stars, as was shown by \citet{MCdonald2019}, who statistically
studied the radius distribution of sub-Neptune-mass planets in
relation to the amount of X-ray flux they received from the host
stars integrated over their lifetime. However, this yields no
information on individual systems.}

Recently, we developed a framework that used this connection to
extract the XUV evolution of a star from the current properties of
orbiting planets. The framework uses a Bayesian scheme to track
the evolution of a{ primordial hydrogen-dominated} planetary
atmosphere (i.e. it tracks the evolution of the planetary radius)
as a function of the stellar XUV flux evolution history
considering the system parameters (planetary mass, orbital
separation, stellar mass, current stellar rotation period, and age
of the system), their uncertainties, and the currently observed
planetary radius \citep{kubyshkina2019}.

Within the framework of our model, the ideal objects to study are
close-in planets of sub-Neptune to Neptune masses because they are
highly affected by atmospheric escape, but still retain a
significant fraction of their primordial hydrogen-dominated
atmospheres. \citet{kubyshkina2019} tested the framework for a
wide range of system parameters and then applied it to the
HD\,3167 and K2-32 planetary systems, each containing one planet
with the appropriate characteristics for our analysis. However,
the tests carried out and presented in that work indicated that
the constraint on the evolution of the stellar XUV flux may
improve significantly if the analysis is carried out
simultaneously considering multiple {planets in} the same system.

\begin{table*}
\caption{Adopted planet parameters for the Kepler-11 system.}
\label{tab::alles} \centering
\begin{tabular}{c|c|c|c|c|c}
  \hline
  planet & $d_0$ [AU] & \Mpl [\Me]& \Rpl [\Re] & $\rho$ [g\,cm$^{-3}$] & $f_{\rm at,now}$ [$\%$]\\
  \hline
  b & $0.091\pm0.001$ (a) & $2.78^{+0.64}_{-0.66}$ (a) & $1.83^{+0.07}_{-0.04}$ (a) & $2.45^{+0.63}_{-0.66}$ (a) & $0.043^{+0.035}_{-0.026}$ (b) \\
    &                     & & & & $0.006^{+0.01}_{-0.005}$ (c)\\
    \hline
  c & $0.107\pm0.001$ (a) & $5.00^{+1.30}_{-1.35}$ (a) & $2.89^{+0.12}_{-0.04}$ (a) & $1.11^{+0.32}_{-0.32}$ (a) & $1.52^{+0.34}_{-0.30}$ (b)\\
    &                     & & & & $0.95^{+0.44}_{-0.41}$ (c)\\
    \hline
  d & $0.155\pm0.001$ (a) & $8.13^{+0.67}_{-0.66}$ (a) & $3.21^{+0.12}_{-0.04}$ (a) & $1.33^{+0.14}_{-0.15}$ (a) & $3.33^{+0.28}_{-0.26}$ (b) \\
    &                     & & & & $2.72^{+0.29}_{-0.27}$ (c) \\
    \hline
  e & $0.195\pm0.002$ (a) & $9.48^{+0.86}_{-0.88}$ (a) & $4.26^{+0.16}_{-0.07}$ (a) & $0.66^{+0.08}_{-0.09}$ (a) & $9.39^{+0.78}_{-0.73}$ (b) \\
    &                     & & & & $6.97^{+0.57}_{-0.55}$ (c) \\
    \hline
  f & $0.250\pm0.002$ (a) & $2.53^{+0.49}_{-0.45}$ (a) & $2.54^{+0.10}_{-0.04}$ (a) & $0.83^{+0.18}_{-0.16}$ (a) & $1.21^{+0.21}_{-0.18}$ (b) \\
    &                     & & & & $0.91^{+0.30}_{-0.26}$ (c)\\
    \hline
  g & $0.466\pm0.004$ (a) & $<27$ (a) & $3.33^{+0.26}_{-0.09}$ (a) & { $<4.5$} (a) & -- \\
    &                     & & & & \\
  \hline
\end{tabular}
\tablefoot{$f_{\rm at,now}$ is the current atmospheric mass
fraction; (a) -- \citet{bedell2017}; (b) -- based on
\citet{lopez2014}; (c) -- based on \citet{johnstone2015}}
\end{table*}

With six rather low-density sub-Neptune-like planets, the
Kepler-11 system appears to be an ideal target for {applying} our
framework. The Kepler-11 system is composed of six closely packed
planets within 0.5\,AU, with planetary radii in the range of 1.8
to 4.3\,\Re\ but with densities that are significantly lower than
those of {the terrestrial planets and ice giants} in the Solar
System. From an analysis of the first 16 months of Kepler
photometry, \citet{lis11} derived the planetary radii from
transits and masses from transit timing variations, concluding
that the five innermost Kepler-11 planets have bulk densities
between 0.5 and 3.0\,g\,cm$^{-3}$. \citet{lis13} revised the
previous analyses {by employing} 40 months of Kepler photometry
and obtained slightly smaller planetary radii and significantly
lower masses for the two innermost planets. This reduced the range
of bulk densities to 0.6--1.7\,g\,cm$^{-3}$. In both studies, the
stellar properties were obtained from the analysis of one
\citep{lis11} or two \citep{lis13} Keck\,I spectra that were
analysed with spectroscopy made easy
\citep{valenti1996,valenti2005}. From comparisons of the
spectroscopic analysis results with stellar evolutionary tracks,
they finally adopted a stellar mass of $0.961\pm0.025$\,\Msun\ and
an age of 8.5$^{+1.1}_{-1.4}$\,Gyr.

\citet{bedell2017} revised the stellar properties using 22 newly
obtained Keck spectra, deriving a stellar mass of
1.042$\pm$0.005\,\Msun\ and a system age of 3.2$\pm$1.5\,Gyr,
indicating that Kepler-11 is very similar to the Sun. In addition
to comparisons with stellar evolutionary tracks,
\citet{bedell2017} inferred the age of the system by employing
gyrochronology (obtaining an age of 3--3.4\,Gyr), lithium
abundance (obtaining an age of 3.5$\pm$1.0\,Gyr), the
yttrium-to-magnesium [Y/Mg] abundance ratio (obtaining an age of
4.0$\pm$0.7\,Gyr), and stellar activity (obtaining an age of
$\sim$1.7\,Gyr). All these estimates are well below those of
\citet{lis13}. By changing the stellar properties,
\citet{bedell2017} also revised the planetary properties and
obtained somewhat different masses compared to previous estimates,
but the bulk densities changed only slightly. We list the
planetary masses, radii, and bulk densities obtained by
\citet{bedell2017} in Table~\ref{tab::alles}.

We present here the results obtained from simultaneously modelling
the atmospheric evolution of all Kepler-11 planets considering the
more recent parameters given by \citet{bedell2017}. We also
describe here a few upgrades to the modelling framework described
by \citet{kubyshkina2019}. We give constraints on the evolution of
the XUV emission of the host star and on the initial atmospheric
mass fractions of the Kepler-11 planets. This paper is organised
as follows. Section~\ref{sec::model} gives a brief description of
the modelling framework and of its upgrades with respect to what
is described in \citet{kubyshkina2019}.
Sections~\ref{sec::results} and \ref{sec::discussion} present the
results and their discussion, and Section~\ref{sec::conclusions}
draws the conclusions of this work.

\section{Modelling approach}\label{sec::model}
The modelling framework we used for the analysis of the Kepler-11
system was developed and thoroughly described in
\citet{kubyshkina2019}, but with a few improvements that we detail
here. The modelling approach combines three main ingredients that
are necessary to study the evolution of a planetary atmosphere: a
model of the stellar flux evolutionary track, a model that relates
planetary parameters and atmospheric mass, and a model for
computing atmospheric escape rates.

The amount of XUV flux emitted by a late-type star is tightly
related to the stellar rotation rate, and the evolutionary tracks
for the stellar rotation rate of young ($\lesssim$2\,Gyr)
late-type stars may follow different paths ranging from slow
(rotation period longer than 8\,days) to fast (rotation period
shorter than 3\,days) rotation \citep{johnstone2015rot,tu2015}. To
account for this, we modelled the rotation period (in days) as a
power law in $\tau$ (age), normalised such that the rotation
period at the present age ($T_{\rm age}$) is consistent with the
observed stellar rotation period ($P_{\rm rot}^{\rm now}$). We
obtained

\begin{equation}
\label{eq:Trot1} P_{\rm rot} =
    \begin{cases}
    P_{\rm rot}^{\rm now} \left(\frac{\tau}{T_{\rm age}}\right)^{0.566}\,, & \text{$\tau\geq 2$~Gyr}\\
    P_{\rm rot}^{\rm now} \left(\frac{2 {\rm Gyr}}{T_{\rm age} {\rm [Gyr]}}\right)^{0.566}\, \left(\frac{\tau {\rm [Gyr]}}{2 {\rm Gyr}}\right)^{x}\,,& \text{$\tau<2$~Gyr}
    \end{cases}
\end{equation}
where the exponent $x$ is a positive value that typically ranges
between 0 and $\sim$2 and controls the stellar rotation period at
ages younger than 2\,Gyr. After 2\,Gyr, the different paths
converge into one, for which we take the empirical power law given
by \citet{mamajek2008}.

From the rotation period at each moment in time, we derived the
stellar X-ray and EUV luminosities following \citet{pizzolato2003}
and \citet{wright2011}, which relate rotation rates and stellar
masses to X-ray luminosity, including saturation effects for the
fastest rotators, and \citet{sanz2011}, which relate X-ray and EUV
fluxes. To account for variations in stellar bolometric
luminosity, hence of the planetary equilibrium temperature, with
time, we used the MESA isochrones and stellar tracks
\citep[MIST,][]{paxton2018} model grid.

To estimate the planetary mass-loss rate at each moment in time
from the stellar flux and planetary parameters (mass \Mpl, radius
\Rpl, equilibrium temperature \Teq, and orbital separation $d_0$),
we used the analytic formulae given by
\citet{kubyshkina2018approx}, which have been derived fitting the
results of a large grid of one-dimensional hydrodynamic
upper-atmosphere models for planets hosting a hydrogen-dominated
atmosphere \citep{kubyshkina2018grid}. {Our hydrodynamic approach
for modelling atmospheric escape considers only thermal escape and
accounts for ionisation, dissociation, recombination, and
L$\alpha$- and H$_3^+$-cooling. It therefore ignores ion-pickup
interactions with the stellar wind. However, the inclusion of
these non-thermal processes would not affect our results. This is
shown by comparing the thermal and non-thermal mass-loss rates:
\citet{kislyakova2014} estimated that mass-loss rates from
non-thermal processes for the Kepler-11 planets are between
$1.1\times 10^7 {\rm g~s^{-1}}$ and $6.8\times 10^7 {\rm
g~s^{-1}}$, while the thermal mass-loss rates at the present time
range between $10^9$ and $10^{11} {\rm g~s^{-1}}$, and were even
higher during the early stages of evolution.}

To estimate the atmospheric mass of a planet, $M_{\rm atm}$, as a
function of planetary parameters, \citet{kubyshkina2019} employed
the results of the model presented by \citet{johnstone2015}, who
pre-computed a grid of atmospheric masses for the range of
planetary parameters covered by
\citet{kubyshkina2018grid,kubyshkina2018approx}, among which we
further interpolated during an evolution run. However, a series of
works studying the Kepler-11 system
\citep[including][]{lis11,lis13,ikoma2012} used the approximation
given by \citet{lopez2013,lopez2014}. For this reason, to enable
comparisons between our and previous results, particularly with
respect to the initial atmospheric mass fraction $f_{\rm at,0}$,
we here employ the approximation of \citet{lopez2014} as the prime
tool for deriving atmospheric mass fractions from the system
parameters. However, we compared the two approaches for the
specific case of the Kepler-11 system and obtained that the
approximation of \citet{lopez2014} leads to a slightly higher
atmospheric mass than the model by \citet{johnstone2015}. The
differences increase with planetary mass and reach about 7\% for
planet e. We further tested the dependence of the final results on
this choice and found that the differences are not significant.
{For both approaches, the core composition has only a minor
influence on the result.}

In our previous works we modelled the planetary atmospheric
evolution \citep{kubyshkina2018grid,kubyshkina2019} and considered
an initial planetary radius (and therefore $f_{\rm at,0}$)
corresponding to a value of the restricted Jeans escape parameter
$\Lambda$ equal to 5, where $\Lambda$ is the Jeans escape
parameter at the position of the planetary radius and for a
temperature equal to \Teq, for a hydrogen-dominated atmosphere
\citep{jeans1925,fossati2017}. As we showed in
\citet{kubyshkina2019}, this approximation does not influence the
result for the majority of planets, but some of the planets in the
Kepler-11 system (d, e, and likely g) are an exception. We
therefore further set the initial atmospheric mass fraction of all
planets as a free parameter here.

Combining the ingredients described above enables computing
planetary atmospheric evolutionary tracks for any set of input
parameters. We assumed that the stellar mass remains constant with
time and ignored the contribution of gravitational contraction and
radioactive decay on \Teq\ during the first phases of evolution.
We furthermore assumed that each planet accreted a
hydrogen-dominated atmosphere from the protoplanetary nebula. The
starting time of the simulations was set {to 1\,Myr, but we ran
tests for starting ages ranging between 0.3 and 10\,Myr (see
Section~\ref{sec::results}). The typical lifetimes of
protoplanetary discs would be around 5\,Myr \citep{mamajek2009},
but we focused on a younger age to enable comparison of our
results with accretion models (see
Section~\ref{sec::discussion}).}

The analysis in this paper relies on the assumption that the
orbital semi-major axes of the Kepler-11 planets have remained
roughly constant since the dispersal of the gas disc. We now
justify this assumption by arguing that only a very minor change
in semi-major axes of a few percent can have taken place: a strong
instability in the system's past is unlikely, but tidal forces may
have caused a minor change in semi-major axes.

When the parameters of \citet{bedell2017} are used, the inner five
planets are separated by {8.0, 15.7, 8.9, and 11.1} mutual Hill
radii, while orbital eccentricities are low ($<0.05$) and the
system is coplanar to within $1^\circ$. This compact, dynamically
cold configuration likely rules out any significant orbital
changes since the epoch of planet formation, which must have
occurred within the gas disc for the planets to have acquired
their significant hydrogen envelopes. The high multiplicity, low
eccentricities and inclinations, and near-resonant period ratios
are characteristic of super-Earth systems that remain stable after
formation in the gas disc \citep{izidoro2019,lambrechts2019}.
{\citet{lis13} showed that assigning high eccentricities resulted
in orbital instability in $<1$\,Myr, which additionally favours a
dynamically cold configuration (low eccentricity and inclination)
that has not undergone a past instability.}

Although the planets have likely not undergone any major orbital
changes since their formation, small changes in their semi-major
axes could have occurred as a result of tidal deformation of the
planets. Using the constant-$Q$ tidal model of \citet{jackson2008}
and assuming a planetary tidal quality factor
$Q^\prime_\mathrm{pl}$ of $100$,  we find a timescale for tidal
eccentricity damping of the inner planet of $\sim$600\,Myr, so
that some tidal evolution in the system is certainly possible.
Kepler-11b and~c are just wide of the 5:4 mean motion resonance,
and differential tidal dissipation of neighbouring resonant
planets has been invoked to explain why many Kepler systems have
orbital period ratios just wide of exact integer ratios
\citep[e.g.,][]{delisle2014a,delisle2014b}. In such scenarios, the
orbital semi-major axes change by only a few percent because the
damping ceases when the planets move away from the resonance and
the associated eccentricity forcing weakens. Such a small change
in orbital semi-major axes will result in a similarly small change
in the incident XUV flux on the planets, much smaller than the
range of fluxes corresponding to the range of stellar rotation
histories considered in this paper.

At each step of the evolution, we first computed the mass-loss
rate based on the stellar flux and system parameters, which we
used to update the atmospheric mass fraction and the planetary
radius. We adjusted the time step such that the change in
atmospheric mass $M_{\rm atm}$ is lower than 5\%. Finally, we
applied a Bayesian approach to constrain the planetary initial
mass fractions and the evolutionary tracks of the stellar XUV
luminosity by fitting the currently observed planetary radii. To
this end, we combined the planetary evolution model with the
open-source Markov chain Monte Carlo (MCMC) algorithm of
\citet{cubillos2017} to compute the posterior distributions for
$f_{\rm at,0}$ for each planet, the stellar rotation rate, and the
considered system parameters.

For each MCMC run, we let the planetary masses, age of the system,
present-time rotation period, orbital separations, and stellar
mass be free parameters, with Gaussian-like priors according to
the measured values and uncertainties. {The present-time rotation
period of the star was defined using the surface rotation velocity
and set as $24.62^{+3.47}_{-2.27}$ days \citep{bedell2017}. The
goodness of fit was quantified using chi-squared statistics.} For
the unknown parameters (initial atmospheric masses and the power
law $x$ in Equation~(\ref{eq:Trot1}), which sets the evolution of
the stellar XUV luminosity), we set a uniform prior within the
following ranges. For $x$, we considered a range from 0 to 2,
where 0 corresponds to the stellar rotation period remaining
constant before 2\,Gyr and 2 corresponds to periods shorter than
0.3\,days for the parameters of Kepler-11. For the atmospheric
mass fractions of the five inner planets, we set uniform priors
{from 0 to 50\%}, while for the possibly high-mass planet g, we
set a uniform prior between 0 and 100\%. Increasing the upper
limit for the atmospheric mass fractions does not significantly
affect the results. As the degree of the power law $x$ given in
Equation~(\ref{eq:Trot1}) is not intuitively clear, we provide the
results in terms of the stellar rotation period at the age of
150\,Myr. This particular age was chosen because it allows
comparison with the distribution of stellar rotation periods
measured in young open clusters \citep{johnstone2015rot}.

\begin{figure*}
  \includegraphics[width=\hsize]{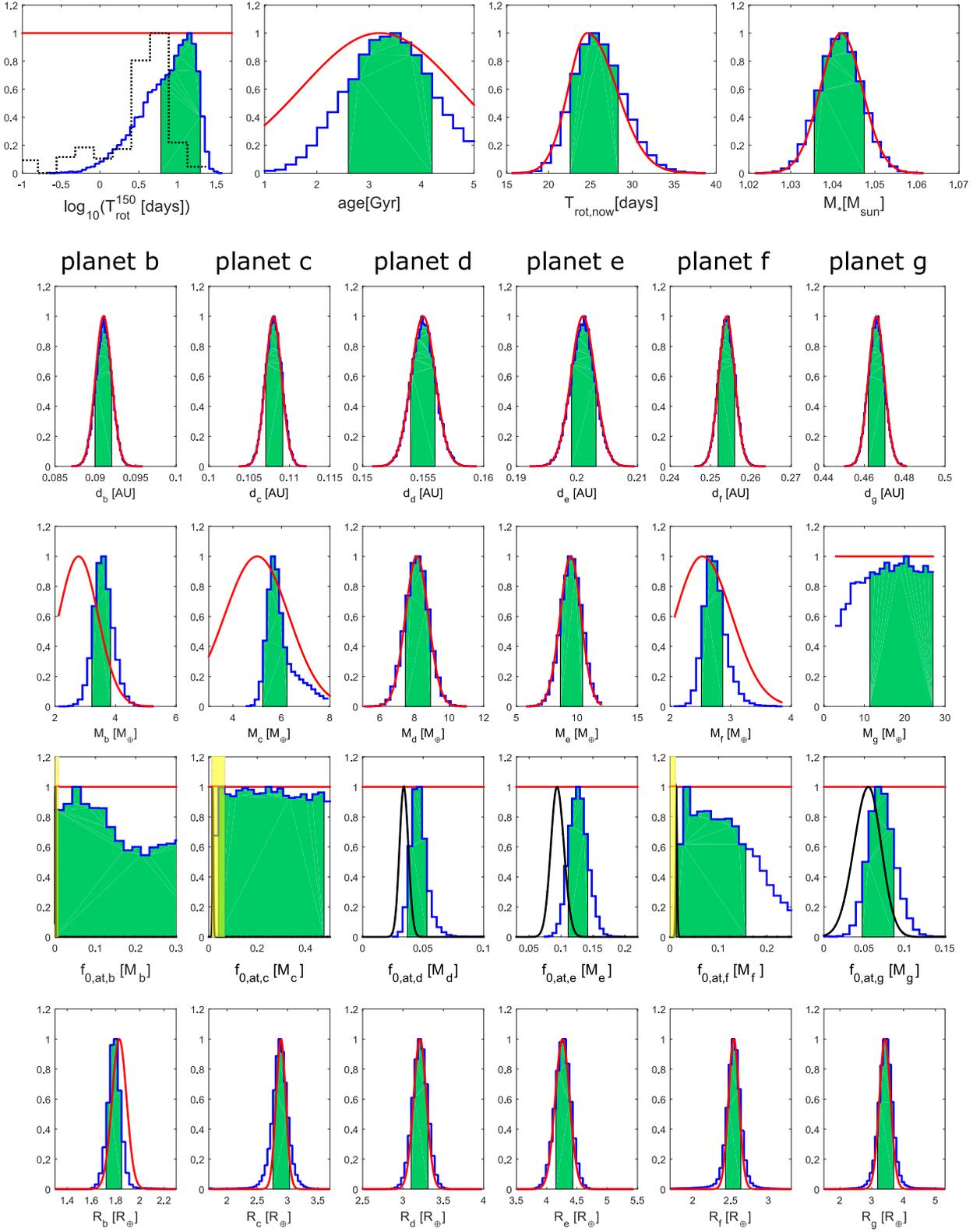}\\
  \caption{Posterior probabilities for the considered Kepler-11 system parameters. Top row -- stellar parameters, from left to the right: rotation period
  at an age of 150\,Myr, age of the system, present time stellar rotation period, and stellar mass. Second row -- planet orbital distance ($d$).
  Third row -- planetary mass. Forth row -- initial atmospheric mass fraction {(i.e., at an age of 1\,Myr)}. {Bottom row -- planetary radius.}
  The blue solid lines indicate the posterior probabilities, the green shaded areas correspond to the 68\% HPD credible intervals, and the red solid lines are the priors.
  The dashed black line in the top-left panel shows the distribution measured {for solar mass members} of $\approx$150\,Myr-old open clusters \citep{johnstone2015rot}.
  {The black solid lines in the 4th row illustrate the present time atmospheric mass fractions obtained using the approximation given by \citet[][solid line]{lopez2014} for the posteriors given by MCMC.}
  The yellow shaded areas in the 4th row are the initial atmospheric mass fractions given by the accretion models of \citet{ikoma2012}
  for a disk temperature of 550\,K {at 0.25\,AU, the location of planet f} and a disk dispersal time of 1\,Myr (see text).}\label{fig::results::bed}
\end{figure*}

%
\section{Results}\label{sec::results}
As mentioned above, we considered the system parameters given by
\citet{bedell2017}. We obtained a rotation period at 150\,Myr
ranging between {5.7 and 19.7\,days}
(Figure~\ref{fig::results::bed}, top left panel), which
corresponds to the long-period wing of the distribution obtained
for stars in $\approx$150\,Myr old open clusters
\citep{johnstone2015rot}. In terms of X-ray luminosity, this range
corresponds to values of about 1 and 10 times the X-ray luminosity
of the present Sun. {The posterior distributions of the system
parameters are consistent with the priors, except for the mass of
planet b, for which the posterior peaks at the 1$\sigma$ upper
boundary of the prior. The planet lies very close to the host
star, making it subject to powerful escape. For this reason, to
avoid complete atmospheric escape, the framework tends to increase
the mass of the planet because planet b is likely to still host a
shallow hydrogen-dominated atmosphere (see
Table~\ref{tab::alles}), as discussed in detail in
\citet{kubyshkina2019}.}

Except for planet g, for which the prior mass is unconstrained,
the mass posteriors match the prior distributions well for the
heavier planets in the system (i.e. d and e), while the posteriors
are significantly narrower than the priors for the lighter planets
(b, c, and f). The 68$\%$ higher posterior density (HPD) credible
intervals are 3.2-4.0\,\Me, 5.2-6.2\,\Me, and 2.5-2.9\,\Me\ for
planets b, c, and f, respectively. For planet g, the posterior
distribution decreases steeply to lower masses with a minimum of
about $50\%$, and is almost uniform above 11\,\Me. {Although
formally the lower boundary of $68\%$ HPD interval lies at
11\,\Me\, , the shape of the probability distribution function
does not enable us to place any firm constraint on the planetary
mass.}

The $f_{\rm at,0}$ posterior distributions (bottom row of
Figure~\ref{fig::results::bed}) are close to uniform for the two
innermost planets, but are well constrained for planets d, e, f,
and g. {For planet b, the distribution has a broad maximum below
an $f_{\rm{at,0}}$ of about 15\% (with a peak at 5\%), and it
remains practically uniform above this value. For planet c, we
obtain a lower $f_{\rm{at,0}}$ limit of 4.0\%. For planet f, the
least massive in the system, the posterior probability
distribution is nearly flat between $f_{\rm{at,0}}$ values of 4
and 13\%, decreases steeply above this, has an isolated peak at
3\%, and is cut abruptly below 1\%. When the lifetime of the
protoplanetary disc is increased above 3\,Myr, the peak at 3\%
becomes less pronounced and the upper limit of 68\% credible
interval decreases to 10\%. The posterior distributions for
planets d and e have a Gaussian-like shape, and the HDP credible
intervals lie in the 3.7--5.3\% and 11.1--14.4\% ranges,
respectively.} From the $f_{\rm at,0}$ posterior distributions we
conclude that the atmospheres of the two high-mass planets (d and
e) likely remained nearly unaffected by atmospheric escape, while
the low-mass planets b, c, and f have lost a significant amount of
hydrogen, up to more than 10 times their present-day atmospheres.

{It may look surprising that although the mass of planet g is
unconstrained, the $f_{\rm at,0}$ posterior probability
distribution is quite narrow. The reason is that a small change in
the fraction of hydrogen leads to a large change in planetary
radius (well constrained by the observations) but to a small
change in the planetary mass. To illustrate this, we considered a
planet (orbiting at the position of planet g) with a mass of 10
\Me and an atmospheric mass fraction of $5\%$. By increasing
$f_{\rm at}$ by just $1\%,$ the planetary radius increases by
about 0.2 \Re. Reaching {the same change in atmospheric mass
fraction} by changing the planetary mass requires an increase in
$M_{\rm pl}$ by 5 \Me.}

\begin{figure}
  \includegraphics[width=\hsize]{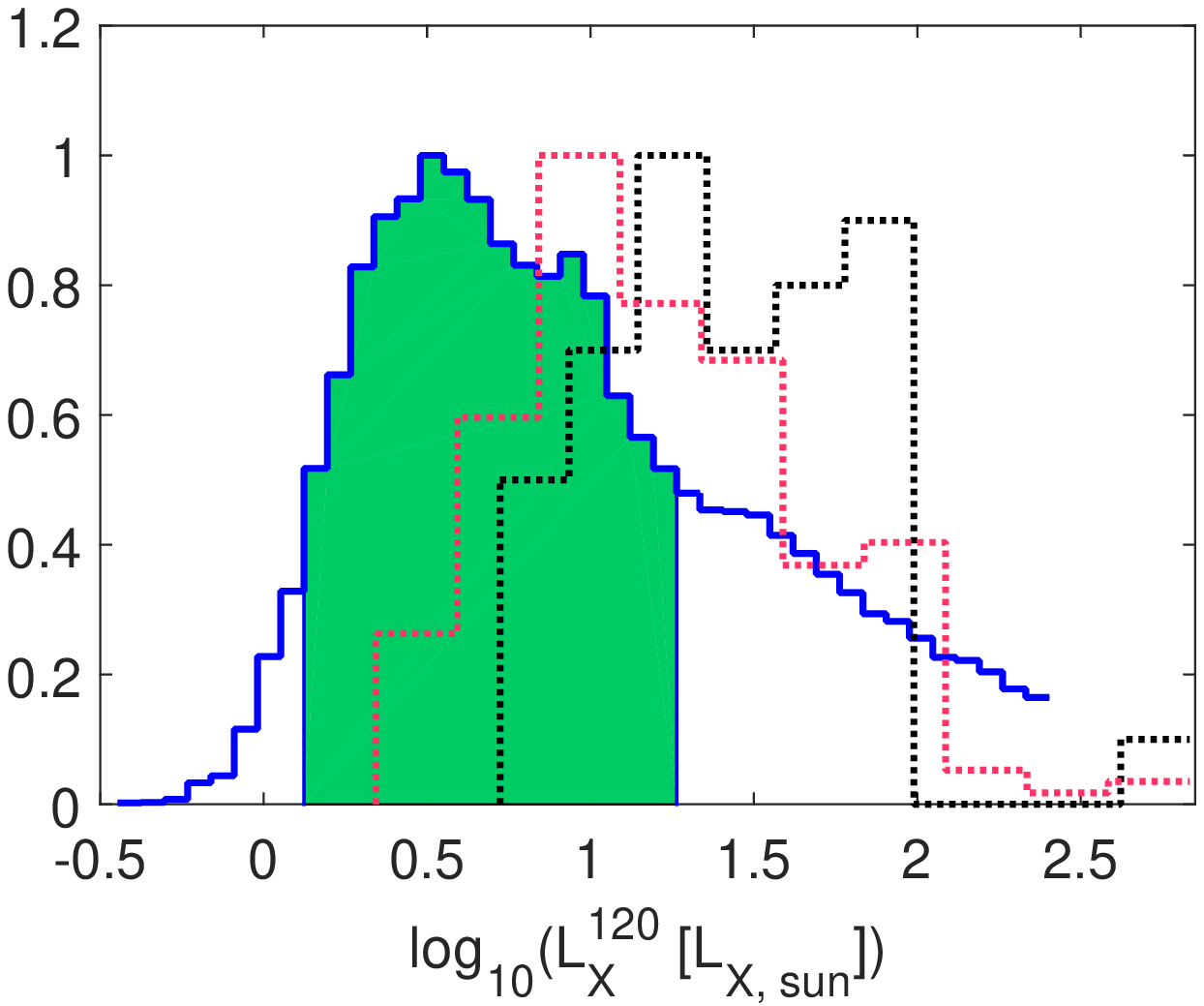}\\
  \caption{{Posterior probability distribution for $L_{\rm x}$ for Kepler-11 at an
  age of 120 Myr (solid blue line; 68\% confidence interval in green) in comparison to the $L_{\rm x}$ distribution obtained for stars of similar spectral
  type in NGC 2516 (black dotted line; 48 stars) and for all detected stars in the same
  cluster (red dotted line; 239 stars; \citet{pillitteri2006}).}}\label{fig::lx}
\end{figure}

{We used the empirical conversion between the rotational period of
the star and its X-ray luminosity derived by \citet{wright2011}.
However, this relation may not apply equivalently well for stars
of different spectral type. To verify this, we collected X-ray
measurements obtained for stellar members of the young (120 Myr)
open cluster NGC 2516 \citep{pillitteri2006} and compare them to
our results in Figure \ref{fig::lx}. This plot indicates that
Kepler-11 was less active than the average stars of similar
spectral type, which agrees with our results based on the stellar
rotation period.}

\section{Discussion}\label{sec::discussion}
Based on the system parameters given by \citet{lis11},
\citet{ikoma2012} studied atmospheric accretion within the nebula
for the five inner planets assuming a temperature of the
protoplanetary disc ranging between 200 and 940\,K and
disc-dispersal ages ranging between 0.01 and 10\,Myr. Therefore,
the work of \citet{ikoma2012} provides a way to use the planetary
atmospheric mass fractions derived above to infer the properties
of the protoplanetary disc.

The model of \citet{ikoma2012} assumes a grainless atmosphere with
a solar H/He abundance on top of a rocky body\footnote{The density
of the rocky core has only a minor effect on the result.} {in
pressure balance with the protoplanetary disc at either the Bondi
radius or the Hill radius, whichever is smaller,} and heated by
the rocky core from below. The authors assumed that the planet is
hot at the beginning of the simulation but has no further energy
supply, so that the rocky body cools rapidly, on a timescale much
shorter than the atmospheric accretion timescale. The
protoplanetary disc dissipates concurrently with the growth of the
atmosphere until full dissipation. The decrease in disc density
results in atmospheric cooling. Furthermore, during disc
dissipation, the atmospheres of planets whose masses are too low
to undergo runaway accretion experience significant erosion due to
the expansion of the atmosphere as a consequence of the decreasing
external pressure provided by the disc
\citep{steokl2015,owen2017}. Because of their low masses, this
also applies to the planets in the Kepler-11 system.

A direct combination of the results of \citet{ikoma2012} with our
results is not strictly possible because of differences in the
considered planetary and stellar masses; but the masses of planets
d and e did not change significantly between \citet{lis11}, which
was considered by \citet{ikoma2012}, and \citet{bedell2017}, which
we considered here. For this reason, we searched for the
parameters of the protoplanetary disc in the work of
\citet{ikoma2012} that best fit the mass and $f_{\rm at,0}$
posterior distributions of planets d and e. We obtained a disc
dispersal time of 1\,Myr and {a temperature of 550\,K at the
position of planet f}. After then fixing the disc dispersal time
and temperature with the results for planets d and e, we
determined the range of atmospheric mass fractions predicted by
\citet{ikoma2012} for planets b, c, and f corresponding to the
range of planetary masses given by our probability distributions
for these three planets. {We obtained $f_{\rm at,0}$ values in the
following ranges: below $\sim$0.5\% for the planet b,
$\sim$1.5--6.5\% for the planet c, and below 1\% for the planet f}
(see yellow shaded areas in Figure~\ref{fig::results::bed}).
{These ranges lie close to the lower boundary of our estimates,
except for planet f, where the range is practically below the
lower boundary of our HPD credible interval, and it partly lies
even below the present-time atmospheric mass fraction.} This low
value of an initial atmospheric mass fraction seems unlikely and
might suggest that the protoplanetary disc may have survived
longer than 1\,Myr. These results indicate that our analysis has
the potential of providing important constraints to planet
formation models and calls for dedicated formation models that are
more consistent with our input parameters and assumptions.

{For planet g, we cannot place a constraint on the planetary mass.
However, the same accretion model of \citet{ikoma2012} that fits
other planets in the system predicts an initial atmospheric mass
fraction above 70\% for the mass range above $10$\,\Me\ and
position of planet g. This estimate is largely higher than what
the planet could possibly lose through atmospheric escape
according to the results of our modelling. The initial mass of the
atmosphere of planet g estimated in the present work (0.05-0.09
${\rm M_g}$) would indicate, according to the accretion model of
\citet{ikoma2012}, a planetary mass range between 3 and 4 Earth
masses. Although this reasoning cannot be regarded as a strict
constraint, it may indicate that the mass of planet g is
significantly lower than the given upper limit.}

\begin{figure*}[p]
  \includegraphics[width=\hsize]{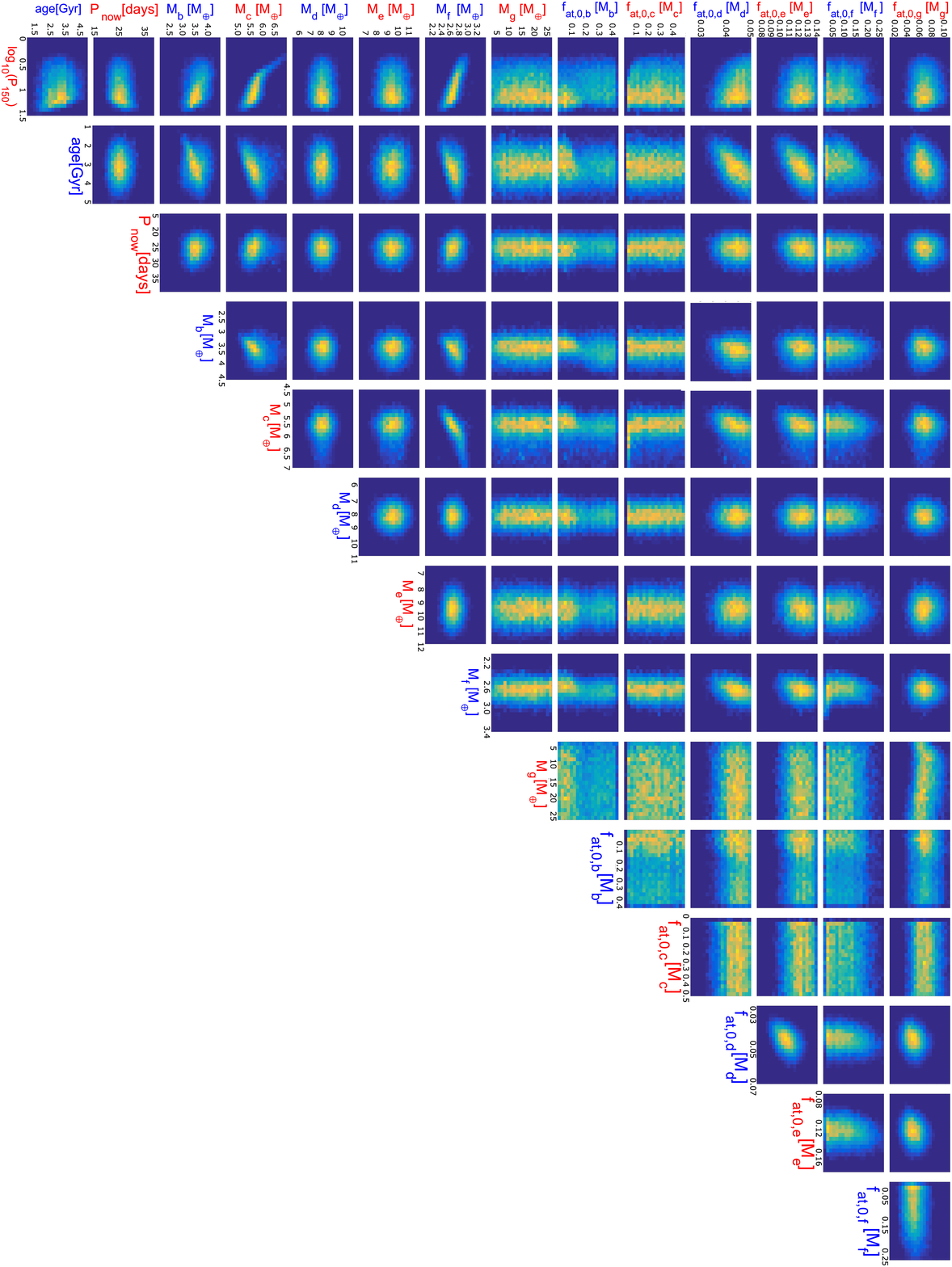}\\
  \caption{Pair-wise distributions of the most relevant system parameters.}\label{fig:results:corner}
\end{figure*}

Figure~\ref{fig:results:corner} presents the pair-wise posterior
distributions of the system parameters. In total, each MCMC run
included 22 parameters, but to improve the readability of the
plot, we excluded stellar mass and planetary orbital separations
because they match the priors and do not show any correlation with
other parameters. We searched for correlations in
Figure~\ref{fig:results:corner}. There is a weak correlation
between age of the system and stellar rotation period at 150\,Myr,
$P_{150}$, where a younger star corresponds to a slower rotator.
The masses of planets b, c, and f are also correlated with
$P_{150}$ (the slower the rotator, the lower the mass of the
planet) and age of the system (the younger the system, the lower
the mass of the planet). The initial atmospheric mass fractions of
planets d, e, and g present a weak correlation with age (younger
systems have smaller initial envelopes). The initial atmospheric
mass fractions of planets b and c are instead independent of both
age and stellar rotation period, indicating that the evolution of
these atmospheres is completely set in the first few megayears
during the saturation phase of the stellar XUV emission. These
correlations arise because the strength and extent of atmospheric
escape, hence the evolution of the planetary radius, depends on
the age of the system, planetary mass, and stellar XUV emission,
which is tightly related to the stellar rotation period
(Section~\ref{sec::model}).

{We employed a specific power law to describe the evolution of the
stellar rotation model. Aiming to quantify possible variations in
the result due to the choice of the model, we tested the effect of
changing the age at which the two parts of the power law
\ref{eq:Trot1} merge ($\tau_{\rm merge}$), which we set to be
equal to 2 Gyr, by changing it between 800 Myr and 3 Gyr. We ran
the full simulation setting $\tau_{\rm merge}$ at 800 Myr, 1.5
Gyr, and 3 Gyr (the run with $\tau_{\rm merge}$ equal to 2 Gyr is
the default one) and obtained that the peak of the posterior
distribution of the stellar rotation period at an age of 150 and
550 Myr changes by  $5\%$ at most, while the edges of the $68\%$
HPD vary by up to $10\%$, corresponding to variations in the
stellar rotation period of up to one day. We further tested this
by also changing the assumed time of the disc dispersal and found
that the variations in the posterior distribution for the stellar
rotation period at an age of 150 Myr are largest for shorter
dispersal times (<3 Myr), while for {longer times} the effect is
smaller, particularly for $\tau_{\rm merge}$ longer than 2 Gyr. We
also found that for the planets with the lowest masses (i.e.
planets b, c, and f) variations in the posterior distributions for
planetary mass and initial atmospheric mass fraction are always
lower than $3\%$ and $5\%$, respectively, while they are
negligible for the high-mass planets (i.e. planets d and e). We
find typically small variations probably because most of the
atmosphere escapes in the first few tens of megayears, which is a
short period of time compared to the timescales described by the
two portions of the power law, hence by the specific position of
$\tau_{\rm merge}$.}

\section{Conclusions}\label{sec::conclusions}

The Kepler-11 system is composed of a star with approximately
solar mass and six planets in the range of super-Earths to Neptune
mass, five of which {orbit} within the distance of Mercury to the
Sun. However, {based on their observed masses and radii}, all
Kepler-11 planets seem to still retain hydrogen-dominated
atmospheres. We employed a planetary atmospheric evolution scheme
in a Bayesian framework to derive the evolutionary path of the
stellar rotation, and therefore high-energy emission, and to
constrain the initial atmospheric mass fractions of the six
planets.

Our results indicate that this exotic configuration with six
close-in low-density planets is possible when the star has evolved
as a slow rotator, following an evolution of the rotation period
covered by about 15\% of all stars in the 0.9--1.1\,\Msun\ mass
range. We also found that the initial atmospheric mass fractions
can be well constrained for high-mass and/or distant planets
because for such planets, atmospheric escape is not strong enough
to dramatically affect the atmosphere, even during the early
stages of evolution. Specifically, in the Kepler-11 system, this
is the case for planets d and e, for which we obtain narrow
posterior distributions of the initial atmospheric mass fraction,
which indicate that both planets have {lost about 35\% of their
initial hydrogen atmospheric content. In particular, for planets d
and e, we find that the initial atmospheric mass fraction was
likely to lie within 3.7--5.3\% and 11.1--14.2\%, respectively.
For the outermost planet (planet g), assuming 10\,\Me, which
corresponds to the formal lower mass limit, escape has led to a
total loss of about 15\% of the initial hydrogen atmospheric
content, with an initial atmospheric mass fraction peaking at
about 6.5\%. For the other planets in the system (b, c, and f),
the probability distributions of the initial mass fraction are
significantly broader, but still enable us to place some
constraints. For planet b, we find that the initial atmospheric
mass fraction was likely below 10\%, for planet c we find that it
was higher than 4\%, and for planet f we find that it was between
1 and 15\%.}

We further employed our results for planets d and e in combination
with those of \citet{ikoma2012} to make a first attempt at
constraining the temperature and lifetime of the protoplanetary
disc. We obtained values of 550\,K {at the position of planet f}
and 1\,Myr. Further comparisons for planets b, c, and f suggest
that a disc-dispersal time of 1\,Myr may be underestimated. These
results should be taken with caution because of differences in the
system parameters considered in this work and by
\citet{ikoma2012}.

{We have shown that planetary atmospheric evolution modelling can
be a very powerful tool for constraining both planet formation
processes (i.e. atmospheric accretion) and stellar characteristics
(i.e. evolution of the stellar rotation rate), in particular for
the latter, which is otherwise difficult, if not impossible, to
obtain by other means. In theory, our analysis framework could be
applied to any planet, but it gives meaningful results only for
systems hosting at least one planet currently holding a
hydrogen-dominated atmosphere and for which the planetary mass and
radius and the age of the system are well measured.} Most
(sub-)Neptunes have so far been detected by the Kepler and K2
missions, but the masses and ages of many of these planets are
poorly constrained. The Transiting Exoplanet Survey Satellite,
TESS \citep{ricker2015} and CHaracterising ExOPlanets Satellite,
CHEOPS \citep{broeg2013} missions are expected to greatly increase
the number of systems that can be analysed in this way, hence
giving us the possibility to constrain planet formation and the
evolution of stellar rotation on the basis of a statistically
significant sample of planets.

%

%
\begin{acknowledgements}
We acknowledge the Austrian Forschungsf\"orderungsgesellschaft FFG
project ``TAPAS4CHEOPS'' P853993, the Austrian Science Fund (FWF)
NFN project S11607-N16, and the FWF project P27256-N27. AJM
acknowledges support from the Knut and Alice Wallenberg Foundation
(2014.0017), the Swedish Research Council (2017-04945) and the
Royal Fysiografical Society in Lund. NVE acknowledges ICM SB RAS
project 0356-2017-0005 and the RFBR grants 18-05-00195-a and
16-52-14006 ANF\_a. We thank the anonymous referee for the careful
reading of the manuscript and for the useful comments that have
led to a significant improvement of the paper.
\end{acknowledgements}

\end{document}